\documentclass[aps,floats]{revtex4}
\usepackage{amsmath,amssymb}
\usepackage{graphicx,epsfig}
\begin{document}
\bibliographystyle {plain}

\def\oppropto{\mathop{\propto}} 
\def\opsimeq{\mathop{\simeq}}
\def\opoverderline{\mathop{\overline}}
\def\operarrow{\mathop{\longrightarrow}}
\def\opsim{\mathop{\sim}}

\def\fig#1#2{\includegraphics[height=#1]{#2}}
\def\figx#1#2{\includegraphics[width=#1]{#2}}

%\newcommand{\fig}[2]{\epsfxsize=#1\epsfbox{#2}} \reversemarginpar 

%%%%%%%%%%%%%%%%%%%%%%%%%%%%%%%%%%%%%%%%%%%%%%%%%%%%%%%%%%%%%%%%%%%%%%%%%%%%
\title{ Dynamical Barriers in the Dyson Hierarchical model  \\
via Real Space Renormalization  } 

%%%%%%%%%%%%%%%%%%%%%%%%%%%%%%%%%%%%%%%%%%%%%%%%%%%%%%%%%%%%%%%%%%%%%%%%%%%%

 \author{ C\'ecile Monthus and Thomas Garel }
  \affiliation{ Institut de Physique Th\'{e}orique, CNRS and CEA Saclay,
 91191 Gif-sur-Yvette, France}

\begin{abstract}

The Dyson hierarchical one-dimensional Ising model of parameter $\sigma>0$ contains long-ranged ferromagnetic couplings decaying as $1/r^{1+\sigma}$ in terms of the distance $r$. We study the stochastic dynamics near zero-temperature via the Real Space Renormalization introduced in our previous work (C. Monthus and T. Garel, arxiv:1212.0643) in order to compute explicitly the equilibrium time $t_{eq}(L)$ as a function of the system size $L$. For $\sigma<1$ where the static critical temperature for the ferromagnetic transition is finite $T_c>0$, we obtain that dynamical barriers grow as the power-law:  $\ln t_{eq}(L) \simeq \beta \left( \frac{ 4 J_0 }{3( 2^{1-\sigma}-1)} \right)  L^{1-\sigma}$. For $\sigma=1$ where the static critical temperature vanishes $T_c=0$, we obtain that dynamical barriers grow logarithmically as :  $\ln t_{eq}(L) \simeq \left[ \beta  \left( \frac{ 4 J_0 }{3 \ln 2} \right) -1 \right] \ln L $. We also compute finite contributions to the dynamical barriers that can depend on the choice of transition rates satisfying detailed balance.

\end{abstract}

\maketitle

\section{ Introduction  }

The Dyson hierarchical Ising model  \cite{dyson} 
has been introduced as a ferromagnetic model where the partition function
could be analyzed via exact renormalization.
The hierarchical couplings are chosen to mimic
  effective long-range power-law couplings $J(r) \simeq 1/r^{1+\sigma}$ in one dimension
so that a phase transition with finite $T_c$ is possible in the region $0<\sigma<1$.
This type of hierarchical model has thus attracted a great interest 
in statistical physics, both among mathematicians
\cite{bleher,gallavotti,book,jona} and among physicists \cite{baker,mcguire,Kim,Kim77}.
Note that Dyson hierarchical models have been 
also introduced in the field of quenched disordered models,
in particular for random fields \cite{randomfield,us_aval}, spin-glasses \cite{sgdysonAT,sgdysonHS,sgdysonR},
and for Anderson localization \cite{bovier,molchanov,krit,kuttruf,fyodorov,EBetOG,
fyodorovbis,us_dysonloc}.

In the present paper, we consider the Dyson hierarchical ferromagnetic Ising model
and we focus on the stochastic dynamics satisfying detailed-balance, such as the Glauber dynamics 
\cite{glauber}. We do not consider the coarsening dynamics starting from a random initial condition
(see the review \cite{bray}) but focus instead on the equilibrium time $t_{eq}$ near zero temperature, i.e. the time needed to go from one ground state (where all spins take the value $+1$)
to the opposite ground state (where all spins take the value $-1$).
In a previous work, we have introduce a real-space renormalization procedure to determine 
this equilibrium time $t_{eq}$ \cite{us_rgdyna} as a function of the system size.
Here we solve the corresponding RG flow for the Dyson hierarchical Ising model.

The paper is organized as follows.
In section \ref{sec_statics}, we recall the important properties of the equilibrium 
of the Dyson hierarchical Ising model.
In section \ref{sec_rgdyn}, we study the stochastic dynamics satisfying detailed-balance
via the real space renormalization procedure introduced in \cite{us_rgdyna}.
The explicit solutions of the renormalization flow for the 'simple' dynamics and for the Glauber dynamics
are presented in sections \ref{sec_simple} and \ref{sec_glauber} respectively.
Our conclusions are summarized in section \ref{sec_conclusion}.
Finally in Appendix \ref{sec_barrier}, we discuss
 the link between dynamical barriers and the energy-cost of a single domain wall.

\section{ Reminder on the statics  of the Dyson hierarchical Ising model }

\label{sec_statics}

\subsection{Definition of the Dyson hierarchical Ising model  }

The Dyson hierarchical Ising model is a model of $2^N$ 
 classical spins $S_i =\pm 1$ where each configuration ${\cal C}=(S_1,S_2,..S_{2^N})$ has for energy
\begin{eqnarray}
U_{2^N}(S_1,...,S_{2^N})  = -\sum_{i<j} J_{ij} S_i S_j
 && = - J_0 \left[ S_1S_2+S_3S_4+S_5S_6+S_7S_8+...  \right]
\nonumber \\
 &&  - J_1 \left[ \left(\frac{S_1+S_2}{2} \right)\left(\frac{S_3+S_4}{2} \right)+
\left(\frac{S_5+S_6}{2} \right)\left(\frac{S_7+S_8}{2} \right) + ...  \right]
\nonumber \\
 &&  - J_2 \left[ \left(\frac{S_1+S_2+S_3+S_4}{2^2} \right)
\left(\frac{S_5+S_6+S_7+S_8}{2^2} \right) + ...  \right] - ....
\nonumber \\
 &&  - J_{N-1}  \left(\frac{S_1+..+S_{2^{N-1}}}{2^{N-1}} \right)
\left(\frac{S_{2^{N-1}+1}+...+S_{2^N}}{2^{N-1}} \right)
\label{Udyson}
\end{eqnarray}
where the positive couplings $J_n$ depend on $n$ via
\begin{eqnarray}
J_n = J_0 2^{ (1- \sigma ) n} 
\label{Jn}
\end{eqnarray}

To make the link with the physics of long-range one-dimensional models,
it is convenient to consider that the sites $i$ of the Dyson model
 are displayed on a one-dimensional lattice, with a lattice spacing unity.
Then the site $i=1$ is coupled via the coupling $J_n/(2^n)^2$ to 
each spin of index $2^{n-1} < i \leq 2^n$. At the scaling level, the hierarchical Dyson model
is thus somewhat equivalent to the
following power-law dependence in the real-space distance $L_n=2^n$
\begin{eqnarray}
J^{eff}(L_n) =  \frac{J_n}{(2^{n})^2 } = \frac{J_0}{2^{n (1+\sigma)} } = \frac{J_0}{L_n^{1+\sigma}}
\label{JneffLR}
\end{eqnarray}
The parameter $\sigma$ is thus the important parameter of the model.

The ground state energy where all spins have the same sign 
\begin{eqnarray}
U_{2^N}^{GS} = U_{2^N}(S_1=1,...,S_{2^N}=1) 
 && = - J_0  2^{N-1}   - J_1  2^{N-2}   - J_2  2^{N-3} - ... - J_{N-1} 
\nonumber \\
 &&
=   - J_0 2^{N-1} \ \ \frac{1-2^{- \sigma N} }{1- 2^{- \sigma} }
 = - \frac{J_0}{2} L_N \frac{1-L_N^{- \sigma } }{1- 2^{- \sigma} }
\label{UGSdyson}
\end{eqnarray}
is extensive in the number $L_N=2^N$ of spins in the region
\begin{eqnarray}
\sigma > 0
\label{sigma}
\end{eqnarray}

The energy cost of the configuration where the 
 first $2^{N-1}$ spins are $(-1)$, whereas the other $2^{N-1} $ spins are $(+1)$ 
\begin{eqnarray}
U_{2^N}^{(2^{N-1},2^{N-1})} - U_{2^N}^{GS} = 2 J_{N-1}
 = 2 J_0 2^{ (1- \sigma ) (N-1)}  = J_0 2^{\sigma} L_N^{1-\sigma}
\label{Udysontwohalfs}
\end{eqnarray}
grows with the distance $L_N=2^N$ for $\sigma<1$, remains constant for $\sigma=1$,
and decays for $\sigma>1$.
It has been shown \cite{dyson} that the critical temperature $T_c$ for the ferromagnetic transition is finite
for $\sigma<1$ only
\begin{eqnarray}
T_c(\sigma<1)>0 \nonumber \\
T_c(\sigma \geq 1)=0
\label{Tc}
\end{eqnarray}

For $\sigma> 1 $, note that Eq \ref{Udysontwohalfs}
 is not very physical since the energy cost of a single domain wall in the middle of the sample
becomes exponentially small instead of remaining finite in the usual one-dimensional ferromagnetic chain.
So, in the following, we will consider only the region
\begin{eqnarray}
0<\sigma \leq 1
\label{regionsigma}
\end{eqnarray}

\subsection{ Real-space renormalization for the equilibrium near zero-temperature }

In this article, we will focus on the regime 'near zero temperature' 
\begin{eqnarray}
 T \ll J_0
\label{Tregimetcpos}
\end{eqnarray}
 where the partition function
is dominated by the two ferromagnetic ground states.
Then the real-space renormalization of the partition function becomes very simple.
The consecutive spins $( S_{2i-1},S_{2i})$ that are coupled by $J_0$ 
are grouped into a single renormalized spin $ S^R_{i}= \frac{S_{2i-1}+ S_{2i}}{2}$
corresponding to the ferromagnetic cluster of the two spins
 \begin{eqnarray}
\vert S^R_{i}=1 > && \equiv \vert S_{2i-1}=1 > \vert S_{2i}=1 > 
\nonumber \\
\vert S^R_{i}=-1 > && \equiv \vert S_{2i-1}=-1 > \vert S_{2i}=-1 > 
\label{Sr}
\end{eqnarray}
In terms of these $2^{N-1}$ renormalized spins for $i=1,2,..,2^{N-1}$,
 the energy of Eq. \ref{Udyson} becomes
\begin{eqnarray}
U_{2^N}(S_1,...,S_{2^N}) 
 && = - J_0 2^{N-1}
% \nonumber \\ &&
  - J_1 \left[ S_1^R S_2^R+  S_3^R S_4^R  + ...  \right]
% \nonumber \\  && 
 - J_2 \left[ \left(\frac{S_1^R+S_2^R}{2} \right)
\left(\frac{S_3^R+S_4^R}{2} \right) + ...  \right] - ....
\nonumber \\
 &&  - J_{N-1}  \left(\frac{S_1^R+..+S^R_{2^{N-2}}}{2^{N-2}} \right)
\left(\frac{S^R_{2^{N-2}+1}+...+S^R_{2^{N-1}}}{2^{N-2}} \right)
\nonumber \\
 && =  - J_0 2^{N-1} + 2^{1-\sigma} U_{2^{N-1}}(S_1^R,...,S^R_{2^{N-1}}) 
\label{UdysonR}
\end{eqnarray}

So if one considers the partition function where the ratio $U/T$ enters,
one obtains that Eq. \ref{UdysonR} corresponds to the following renormalization of the temperature upon the elimination of the lowest generation
\begin{eqnarray}
T_R = 2^{\sigma-1} T
\label{TR}
\end{eqnarray}
This is in agreement with Eq. \ref{Tc} : 
for $\sigma<1$, $T_R$ flows towards the attractive fixed point $T=0$;
for $\sigma>1$,  $T_R$ flows away from the unstable fixed point $T_c=0$.

To prepare the following sections,
 it is also convenient to describe the renormalization of the local field.
 In the initial model of Eq. \ref{Udyson},
the local field on spin $S_i$ is defined as
\begin{eqnarray}
B_i = - \frac { \partial U_{2^N}(S_1,...,S_{2^N}) }{ \partial S_i } = \sum_{j \ne i} J_{ij} S_j
\label{bisingle}
\end{eqnarray}
i.e. for instance for the two first spins $i=1,2$
\begin{eqnarray}
B_1 && =  J_0 S_2 + \frac{J_1}{2}\left(\frac{S_3+S_4}{2} \right)
  +\frac{J_2}{2^2} 
\left(\frac{S_5+S_6+S_7+S_8}{2^2} \right) + ...
  + \frac{J_{N-1} }{2^{N-1}} 
\left(\frac{S_{2^{N-1}+1}+...+S_{2^N}}{2^{N-1}} \right) 
\nonumber \\
B_2 && =  J_0 S_1 + \frac{J_1}{2}\left(\frac{S_3+S_4}{2} \right)
  +\frac{J_2}{2^2} 
\left(\frac{S_5+S_6+S_7+S_8}{2^2} \right) + ...
  + \frac{J_{N-1} }{2^{N-1}} 
\left(\frac{S_{2^{N-1}+1}+...+S_{2^N}}{2^{N-1}} \right) 
\label{bisingle1}
\end{eqnarray}
After the renormalization where $(S_1,S_2)$ have been grouped into the renormalized spin 
$S_1^R= (S_1+S_2)/2$ (Eq \ref{Sr}), the renormalized local field $B_1^R$ on $S_1^R$
reads
\begin{eqnarray}
B_1^R &&  = J_1 \left(\frac{S_3+S_4}{2} \right)
  +\frac{J_2}{2} 
\left(\frac{S_5+S_6+S_7+S_8}{2^2} \right) + ...
  + \frac{J_{N-1} }{2^{N-2}} 
\left(\frac{S_{2^{N-1}+1}+...+S_{2^N}}{2^{N-1}} \right)  
\nonumber \\
&& = 2^{1-\sigma} J_0  \left[   S_2^R
  + 2^{-\sigma} \left(\frac{S_3^R+S_4^R}{2} \right) + ...
  +   2^{- \sigma (N-2)}
\left(\frac{S^R_{2^{N-2}+1}+...+S^R_{2^{N-1}}}{2^{N-2}} \right)  \right]
\label{b1R}
\end{eqnarray}
i.e. it is renormalized by the same factor $ 2^{1-\sigma} $ as the couplings as it should.

\section{ Real Space Renormalization for the dynamics of the Dyson model  }

\label{sec_rgdyn}

\subsection{ Stochastic single-spin-flip dynamics satisfying detailed-balance }

\label{sec_dynamics}

We consider the stochastic dynamics generated by the Master Equation for the probability $P_t ({\cal C} ) $ to be in  configuration ${\cal C}$
 at time t
\begin{eqnarray}
\frac{ dP_t \left({\cal C} \right) }{dt}
= \sum_{\cal C '} P_t \left({\cal C}' \right) 
W \left({\cal C}' \to  {\cal C}  \right) 
 -  P_t \left({\cal C} \right)  \sum_{ {\cal C} '} W \left({\cal C} \to  {\cal C}' \right) 
\label{master}
\end{eqnarray}
where the transition rates $ W \left({\cal C} \to  {\cal C}' \right)$ satisfy the detailed balance condition
\begin{eqnarray}
e^{- \beta U({\cal C})}   W \left( \cal C \to \cal C '  \right)
= e^{- \beta U({\cal C '})}   W \left( \cal C' \to \cal C   \right)
\label{detailed}
\end{eqnarray}
For the Ising model of Eq \ref{Udyson}, where the energy is of the form
\begin{eqnarray}
U({\cal C}) = -\sum_{i<j} J_{ij} S_i S_j
\label{Uspin}
\end{eqnarray}
with the local fields (Eq \ref{bisingle})
\begin{eqnarray}
B_k = - \frac { \partial U ({\cal C}) }{ \partial S_k } = \sum_{i \ne k} J_{ki} S_i
\label{bksingle}
\end{eqnarray}
it is natural to consider the following single-spin-flip transition rates for the flip of the spin $S_k$
\begin{eqnarray}
W \left( S_k \to -S_k \right) = G_0(B_k)  e^{ - \beta S_k B_k } 
\label{wgene}
\end{eqnarray}
where the function $G_0(B)$ is an even positive function of $B$ (see \cite{us_rgdyna} for more details)
\begin{eqnarray}
G_0(B)=G_0(-B) >0
\label{Gposieven}
\end{eqnarray}

In the following, we will consider as two important examples the 'simple' dynamics
\begin{eqnarray}
{ \rm Simple \ Dynamics } \ : \ \  G^{simple}_0(B) =1
\label{defsimple}
\end{eqnarray}
and the Glauber dynamics \cite{glauber}
\begin{eqnarray}
{ \rm Glauber \ Dynamics } \ :  \ \  G^{Glauber}_0(B) = \frac{1}{2 \cosh \beta B}
\label{defglauber}
\end{eqnarray}

\subsection{ Mapping onto a quantum Hamiltonian }

As recalled in detail in \cite{us_rgdyna}, the Master equation with the transition rates of Eq. \ref{wgene}
can be mapped via a similarity transformation onto the quantum Hamiltonian
 \cite{felderhof,siggia,kimball,peschel,castelnovo}
\begin{eqnarray}
 H
&& =  \sum_{ k=1 }^{2^N} G_0 \left( \sum_{i \ne k} J_{ik} \sigma^z_i  \right) 
  \left( e^{ - \beta \sigma^z_k \left( \sum_{i \ne k} J_{ik} \sigma^z_i \right) } -   \sigma^x_k \right)
\label{Hquantum}
\end{eqnarray}
for $2^N$ quantum spins described by Pauli matrices, with the following properties.
The ground state energy exactly vanishes $E_0=0$ and represents the thermal equilibrium.
 The smallest non-vanishing energy $E_1(2^N)$
determines the equilibrium time $t_{eq}(2^N)$ (defined as the largest relaxation time of the Master Eq. \ref{master})
\begin{eqnarray}
t_{eq}(2^N) = \frac{1}{E_1(2^N)} 
\label{taueqe1}
\end{eqnarray}

\subsection{ Real Space Renormalization of the associated quantum Hamiltonian  }

In our previous work \cite{us_rgdyna}, we have introduced a real-space renormalization procedure
for quantum Hamiltonian of the form of Eq. \ref{Hquantum}
to determine $E_1$ near zero-temperature. We now describe
the application to the Dyson hierarchical model.
The quantum Hamiltonian associated to the stochastic dynamics of the Dyson hierarchical
Ising model can be rewritten as a sum of elementary operators (Eq \ref{Hquantum})
\begin{eqnarray}
H && = \sum_{i=1}^{2^{N-1} } \left( h_{2i-1} +h_{2i} \right)
\\
h_{2i-1} +h_{2i}   && =  G_0(J_0 \sigma^z_{2i}+\frac{B^R_i}{2})
\left( e^{ - \beta \sigma^z_{2i-1} \left(  J_{0} \sigma^z_{2i} + \frac{B^R_i}{2} \right) }
-   \sigma^x_{2i-1} \right) 
+ G_0(J_0 \sigma^z_{2i-1}+\frac{B^R_i}{2})
\left( e^{ - \beta \sigma^z_{2i} \left(  J_{0} \sigma^z_{2i-1} + \frac{B^R_i}{2} \right) }
-   \sigma^x_{2i} \right) 
\nonumber
\label{Htotsumcluster}
\end{eqnarray}
with $B^R_i$ is the static renormalized local field introduced Eq. \ref{b1R}.

Let us consider the first two spins : as explained in \cite{us_rgdyna},
the sum
\begin{eqnarray}
h_1 +h_2  && =  G_0(J_0 \sigma^z_2+\frac{B_1^R}{2})
\left( e^{ - \beta \sigma^z_1 \left(  J_{0} \sigma^z_2 + \frac{B_1^R}{2} \right) }
-   \sigma^x_1 \right) 
% \nonumber \\ &&
 + G_0(J_0 \sigma^z_1+\frac{B_1^R}{2})
\left( e^{ - \beta \sigma^z_2 \left( J_{0} \sigma^z_1 +\frac{B_1^R}{2} \right) }
-   \sigma^x_2 \right)
\label{Hklocalsum}
\end{eqnarray}
can be renormalized onto the following operator describing the flip of the renormalized spin
$S^R = (S_1+S_2)/2$ (Eq \ref{Sr})
\begin{eqnarray}
h_{(1,2)}^R \equiv G_1^R(B_1^R) 
\left( e^{ - \beta \sigma^z_R B_1^R } -   \sigma^x_R \right) 
\label{Hrgop}
\end{eqnarray}
where the renormalized amplitude $G_1^R(B)$ can be computed from the function $B_0$ via
 \begin{eqnarray}
G_1^R (B) 
&& =  e^{-  \beta  J_{0} } 
 \frac{ 2 G_0( J_0 +\frac{B}{2}) G_0(J_0- \frac{B}{2} )   }
{ e^{\beta \frac{B}{2}} G_0( J_0 + \frac{B}{2} ) + e^{-\beta \frac{B}{2}} G_0( J_0-\frac{B}{2} )   }
\label{grdyson}
\end{eqnarray}
or equivalently using the inverse
 \begin{eqnarray}
\frac{1}{G_1^R (B) }
&& =  \frac{e^{  \beta  J_{0} } }{2} 
\left[  \frac{ e^{\beta \frac{B}{2}}   }
{   G_0( J_0 -\frac{B}{2})   }
 + \frac{  e^{-\beta \frac{B}{2}}    }
{  G_0( J_0 +\frac{B}{2})    } \right]
\label{grdysoninverse}
\end{eqnarray}
In conclusion, the real space renormalization of the quantum Hamiltonian
corresponds for the Dyson model to the renormalization of the function $G(B)$
starting from the initial condition $G_0(B)$ that defines the initial 
dynamics (Eq \ref{Gposieven}).

\subsection{ Equilibrium time $t_{eq}(2^N)$ for  a finite system of $2^N$ spins }

We apply iteratively the previous renormalization rule as follows :

(0) The initial dynamics concerns the Dyson hierarchical model of $2^N$ spins with the couplings $(J_0,..,J_{N-1})$ 
and the function $G_0(B)$ that defines the transition rates of Eq \ref{wgene}.

(1) After 1 RG step, we have a Dyson hierarchical model of $2^{N-1}$ renormalized spins with the couplings
  $(J_1,..,J_{N-1})$ and the function $G_1^R(B) $ obtained by Eq \ref{grdysoninverse}
 \begin{eqnarray}
\frac{1}{ G_1^R(B) } =  
 \frac{e^{  \beta  J_{0} } }{2} \sum_{\epsilon_1=\pm} 
 \frac{  e^{-\beta \epsilon_1 \frac{B}{2}}    }
{  G_0( J_0 +\epsilon_1 \frac{B}{2})    }
\label{gr1pas}
\end{eqnarray}

(2) After 2 RG steps, we have a Dyson hierarchical model of $2^{N-2}$ renormalized spins with the couplings
  $(J_2,..,J_{N-1})$ and the function $G_2^R(B) $ obtained by Eq \ref{grdysoninverse} 
 \begin{eqnarray}
\frac{1}{ G_2^R(B) } =  
 \frac{e^{  \beta  J_{1} } }{2} \sum_{\epsilon_2=\pm} 
 \frac{  e^{-\beta \epsilon_2 \frac{B}{2}}    }
{  G_1( J_1 +\epsilon_2 \frac{B}{2})    }
\label{gr2pas}
\end{eqnarray}

...

$(k)$ After $k$ RG steps, we have a Dyson hierarchical model of $2^{N-k}$ renormalized spins with the couplings
  $(J_k,..,J_{N-1})$ and the function $G_k^R(B) $
 \begin{eqnarray}
\frac{1}{ G_k^R(B) } =  
 \frac{e^{  \beta  J_{k-1} } }{2} \sum_{\epsilon_k=\pm} 
 \frac{  e^{-\beta \epsilon_k \frac{B}{2}}    }
{  G_{k-1}( J_{k-1} +\epsilon_k \frac{B}{2})    }
\label{grkpas}
\end{eqnarray}

..

$(N)$ After $N$ RG steps, only one spin remains, whose flipping is governed by
the function $G_N^R(B) $ 
 \begin{eqnarray}
\frac{1}{ G_N^R(B) } =  
 \frac{e^{  \beta  J_{N-1} } }{2} \sum_{\epsilon_N=\pm} 
 \frac{  e^{-\beta \epsilon_N \frac{B}{2}}    }
{  G_{N-1}( J_{N-1} +\epsilon_N \frac{B}{2})    }
\label{grNpas}
\end{eqnarray}
However since this spin is alone, there is no local field $B=0$,
so we just have to compute the final number $G^{final}_{N}=G_{N}^R(B=0) $ 
 \begin{eqnarray}
\frac{1}{ G^{final}_{N} } &&  = \frac{1}{ G_{N}^R(B=0) } =  
 \frac{e^{  \beta  J_{N-1} } }{ G_{N-1}( J_{N-1} ) } 
\nonumber \\ &&
= e^{  \beta  J_{N-1} } 
 \frac{e^{  \beta  J_{N-2} } }{2} \sum_{\epsilon_{N-1}=\pm} 
 \frac{  e^{-\beta \epsilon_{N-1} \frac{J_{N-1}}{2}}    }
{  G_{N-2}( J_{N-2} +\epsilon_{N-1} \frac{J_{N-1}}{2})    }
\nonumber \\ &&
= e^{  \beta  J_{N-1} } 
 \frac{e^{  \beta  J_{N-2} } }{2}  \frac{e^{  \beta  J_{N-3} } }{2}
\sum_{\epsilon_{N-1}=\pm}  \sum_{\epsilon_{N-2}=\pm}  
 \frac{  e^{-\beta \epsilon_{N-1} \frac{J_{N-1}}{2}} e^{-\beta \epsilon_{N-2} \frac{\left( J_{N-2} +\epsilon_{N-1} \frac{J_{N-1}}{2}\right)}{2}}    }
{  G_{N-3}\left[ J_{N-3} +\epsilon_{N-2} \frac{\left( J_{N-2} +\epsilon_{N-1} \frac{J_{N-1}}{2}\right)}{2} \right]   }
\nonumber \\ &&
= 
 \frac{e^{  \beta \sum_{n=0}^{N-1} J_n } }{2^{N-1}}  
\sum_{\epsilon_{N-1}=\pm}  \sum_{\epsilon_{N-2}=\pm}  .. \sum_{\epsilon_1=\pm} 
 \frac{  e^{- \beta \sum_{n=1}^{N-1} \epsilon_{n } \frac{{\cal B}_{n} }{2}}   }
{  G_{0}\left[ {\cal B}_0 \right]   }
\label{gNfinal}
\end{eqnarray}
in terms of the variables ${\cal B}_n$ that can be computed from the recurrence
 \begin{eqnarray}
{\cal B}_{n-1}=J_{n-1} +\epsilon_{n} \frac{{\cal B}_{n}}{2} 
\label{calBrec}
\end{eqnarray}
with the initial condition
 \begin{eqnarray}
{\cal B}_N=0
\label{calBini}
\end{eqnarray}
The first terms read
 \begin{eqnarray}
{\cal B}_{N-1} && =J_{N-1} 
\nonumber \\ 
{\cal B}_{N-2} && =J_{N-2} + J_{N-1}  \left( \frac{\epsilon_{N-1}}{2} \right)
\label{calBsolufirst}
\end{eqnarray}
and more generally for $0 \leq n \leq N-1$, one obtains
 \begin{eqnarray}
{\cal B}_{n}=J_{n} + \sum_{m=n+1}^{N-1} J_m \prod_{k=n+1}^m \left( \frac{\epsilon_{k}}{2} \right)
\label{calBsolu}
\end{eqnarray}
In the following, we will need 
 \begin{eqnarray}
{\cal B}_{0}(\epsilon_1,..,\epsilon_{N-1}) =J_{0} + \sum_{m=1}^{N-1} J_m \prod_{k=1}^m \left( \frac{\epsilon_{k}}{2} \right)
\label{calBsoluzero}
\end{eqnarray}
and the sum present in the exponential of Eq. \ref{gNfinal} 
 \begin{eqnarray}
\Sigma (\epsilon_1,..,\epsilon_{N-1})  \equiv \sum_{n=1}^{N-1} \epsilon_{n } \frac{{\cal B}_{n} }{2} 
&& =\sum_{n=1}^{N-1} \frac{\epsilon_{n}}{2}
  \left[ J_{n}
 + \sum_{m=n+1}^{N-1} J_m \prod_{k=n+1}^m \left( \frac{\epsilon_{k}}{2} \right) \right]
\nonumber \\ &&
= \sum_{m=1}^{N-1} J_m \left[ \sum_{n=1}^m \prod_{k=n}^m \left( \frac{\epsilon_{k}}{2} \right) \right]
\label{calBsumS}
\end{eqnarray}

In conclusion, the equilibrium time $t_{eq}(2^N)$ 
that can be computed from the final amplitude $G^{final}_{N}$  \cite{us_rgdyna} 
 \begin{eqnarray}
t_{eq}(2^N) = \frac{1}{ 2 G^{final}_{N} }
\label{teqGfinal}
\end{eqnarray}
reads
 \begin{eqnarray}
t_{eq}(2^N)  && 
= \frac{e^{  \beta \sum_{n=0}^{N-1} J_n } }{2^{N}}  
\sum_{\epsilon_{N-1}=\pm}  \sum_{\epsilon_{N-2}=\pm}  .. \sum_{\epsilon_1=\pm} 
 \frac{  e^{- \beta \Sigma (\epsilon_1,..,\epsilon_{N-1})}  }
{  G_{0}\left( {\cal B}_0(\epsilon_1,..,\epsilon_{N-1}) \right)   }
\label{teqfinal}
\end{eqnarray}
where ${\cal B}_0(\epsilon_1,..,\epsilon_{N-1})$ and $\Sigma (\epsilon_1,..,\epsilon_{N-1}) $
are given in Eqs \ref{calBsoluzero} and \ref{calBsumS}.
To determine the leading behavior near zero temperature, we have to distinguish whether 
the initial function $G_0(B)$ does not depend on $\beta$ ( as in the simple dynamics of Eq \ref{defsimple})
or depends on $\beta$ ( as in the Glauber dynamics of Eq \ref{defglauber}).
These two cases are studied respectively in the two following sections.

\section{ Equilibrium time for the 'simple' dynamics }

\label{sec_simple}

\subsection{ Real Space renormalization Solution }

For the initial condition of Eq. \ref{defsimple} that defines the 'simple' dynamics,
the result of Eq \ref{teqfinal} for the equilibrium time becomes
 \begin{eqnarray}
t^{simple}_{eq}(2^N)  && 
 = \frac{e^{  \beta \sum_{n=0}^{N-1} J_n } }{2^{N}}  
\sum_{\epsilon_{N-1}=\pm}  \sum_{\epsilon_{N-2}=\pm}  .. \sum_{\epsilon_1=\pm} 
 e^{- \beta \Sigma (\epsilon_1,..,\epsilon_{N-1})}  
\label{teqsimple}
\end{eqnarray}
where
 \begin{eqnarray}
\Sigma (\epsilon_1,..,\epsilon_{N-1}) 
&& = \sum_{m=1}^{N-1} J_m \left[ \sum_{n=1}^m \prod_{k=n}^m \left( \frac{\epsilon_{k}}{2} \right) \right]
\nonumber \\ &&
=  J_1   \frac{\epsilon_{1}}{2} 
+  J_2 \frac{\epsilon_{2}}{2}  \left[ 1+ \frac{\epsilon_{1}}{2}  \right]
+  J_3 \frac{\epsilon_{3}}{2}  \left[ 1+ \frac{\epsilon_{2}}{2} + \frac{\epsilon_{1} \epsilon_{2}}{2^2}  \right] +...
\nonumber \\ &&
+  J_{N-2} \frac{\epsilon_{N-2}}{2}
 \left[ 1+ \frac{\epsilon_{N-3}}{2} + \frac{\epsilon_{N-3} \epsilon_{N-4}}{2^2} +...
+ \frac{\epsilon_{N-3} \epsilon_{N-4} ... \epsilon_1 }{2^{N-3}}  \right]
\nonumber \\ &&
+  J_{N-1} \frac{\epsilon_{N-1}}{2}
 \left[ 1+ \frac{\epsilon_{N-2}}{2} + \frac{\epsilon_{N-2} \epsilon_{N-3}}{2^2} +...
+ \frac{\epsilon_{N-2} \epsilon_{N-3} ... \epsilon_1 }{2^{N-2}}  \right]
\label{calBsumSexpli}
\end{eqnarray}

\subsection{ Leading term near zero temperature }

\label{sec_simpleleading}

At low temperature, the sum of Eq \ref{teqsimple}
will be dominated by the minimal value of $ \Sigma (\epsilon_1,..,\epsilon_{N-1}) $ of Eq. \ref{calBsumS}.
Since $\epsilon_{N-1} $ appears only in the last line, and since 
the term between the brackets cannot change sign with respect to the first term $1$,
we are led to the choice
 \begin{eqnarray}
\epsilon_{N-1}=-1
\label{epslast}
\end{eqnarray}
and we have now to minimize
 \begin{eqnarray}
\Sigma (\epsilon_1,..,\epsilon_{N-2},\epsilon_{N-1}=-1) 
&& =  J_1   \frac{\epsilon_{1}}{2} 
+  J_2 \frac{\epsilon_{2}}{2}  \left[ 1+ \frac{\epsilon_{1}}{2}  \right]
+  J_3 \frac{\epsilon_{3}}{2}  \left[ 1+ \frac{\epsilon_{2}}{2} + \frac{\epsilon_{1} \epsilon_{2}}{2^2}  \right] +...
\nonumber \\ &&
+   \frac{\epsilon_{N-2}}{2} \left(J_{N-2}-   \frac{ J_{N-1}}{2} \right)
 \left[ 1+ \frac{\epsilon_{N-3}}{2} + \frac{\epsilon_{N-3} \epsilon_{N-4}}{2^2} +...
+ \frac{\epsilon_{N-3} \epsilon_{N-4} ... \epsilon_1 }{2^{N-3}}  \right]
\nonumber \\ &&
-   \frac{ J_{N-1}}{2}
\label{calBsumSexpliepsnm1moins}
\end{eqnarray}
Since from Eq \ref{Jn}
\begin{eqnarray}
J_{N-2}-   \frac{ J_{N-1}}{2} = J_0 2^{ (1- \sigma ) (N-2)} (1 -  2^{-\sigma} )>0
\label{Jndiff}
\end{eqnarray}
we are led to the choice 
 \begin{eqnarray}
\epsilon_{N-2}=-1
\label{epslastnext}
\end{eqnarray}
and so on, so that the minimum is achieved when all $\epsilon_i=-1$.
The corresponding minimum reads using Eq. \ref{Jn}
 \begin{eqnarray}
\Sigma_{min} && = \Sigma (\epsilon_1=-1,\epsilon_2=-1,..,\epsilon_{N-1}=-1) 
 = \sum_{m=1}^{N-1} J_m \left[ \sum_{n=1}^m  \left( - \frac{1}{2} \right)^{m-n+1} \right]
\nonumber \\ &&
= - \frac{1}{3} \sum_{m=1}^{N-1} J_m \left[ 1 - \left( - \frac{1}{2} \right)^m \right]
\label{calBsumSmin}
\end{eqnarray}

so that finally 
 \begin{eqnarray}
t^{simple}_{eq}(2^N)  && 
 \simeq \frac{e^{  \beta \left[ \sum_{n=0}^{N-1} J_n - \Sigma_{min} \right] } }{2^{N}}  
% \nonumber \\ && 
=  
\frac{ 1 }{2^{N}}  \ 
e^{  \beta \left[\displaystyle  J_0+ \sum_{m=1}^{N-1} J_m \left( \frac{4 - \left(- \frac{1}{2} \right)^m}{3} \right) \right] }
\label{teqsimplefinal}
\end{eqnarray}
We may now use Eq \ref{Jn} to compute explicitly
 \begin{eqnarray}
\ln \left[ 2^N t^{simple}_{eq}(2^N) \right] && 
=  \beta \left[  J_0+ \sum_{m=1}^{N-1} J_0 2^{(1-\sigma)m}  \left( \frac{4 - \left(- \frac{1}{2} \right)^m}{3} \right) \right] 
\nonumber \\ && =  
\beta J_0 \left[ 1 + \frac{4}{3} 2^{1-\sigma} \frac{1-2^{(1-\sigma)(N-1)} }{1-2^{1-\sigma}}
+ \frac{1}{3} 2^{-\sigma} \frac{1-(-2^{-\sigma})^{(N-1)} }{1+2^{-\sigma}} \right]
\label{teqsimplefinalexpli}
\end{eqnarray}
This expression is valid for finite $N$ and we may check the first terms
 \begin{eqnarray}
\ln \left[ 2 t^{simple}_{eq}(2) \right] &&   = \beta J_0
\nonumber \\
\ln \left[ 2^2 t^{simple}_{eq}(2^2) \right]  &&  = \beta J_0 \left[ 1+3.2^{-\sigma} \right]
= \beta \left[ J_0+\frac{3}{2} J_1 \right]
\label{teqsimplefirst}
\end{eqnarray}

To obtain the leading behavior for large $N$, we should now specify the value of $\sigma$.

\subsection{ Case  $\sigma<1$ : Power-law barrier  }

For $\sigma<1$, Eq. \ref{teqsimplefinalexpli} becomes in terms of the length $L_N=2^N$
 \begin{eqnarray}
\sigma<1 \ \ : \ \ 
\ln \left[ 2^N t^{simple}_{eq}(L_N=2^N) \right] 
% &&  \simeq \beta J_0 \left[ \frac{4}{3}  \frac{2^{(1-\sigma)N} }{2^{1-\sigma}-1}    +  1 + \frac{4}{3}  \frac{1 }{2^{\si% gma-1}-1} + \frac{1}{3}  \frac{1 }{2^{\sigma}+1} \right]
% \nonumber \\
&& 
= \beta J_0 \left[ \frac{4}{3}  \frac{L_N^{(1-\sigma)} }{2^{1-\sigma}-1}
    +  1 + \frac{4}{3}  \frac{1 }{2^{\sigma-1}-1}
+ \frac{1}{3}  \frac{1 }{2^{\sigma}+1} \right]+O( L_N^{-\sigma})
\label{teqsimplefinalsigmal1}
\end{eqnarray}
i.e. the energy barrier near zero temperature scales as the following power-law of the length $L_N=2^N$
 \begin{eqnarray}
\frac{ \ln \left[ 2^N  t^{simple}_{eq}(L_N=2^N) \right] }{\beta} \simeq  
  \frac{ 4 J_0 }{3( 2^{1-\sigma}-1)}  L_N^{(1-\sigma)} 
\label{teqsimplefinalbarrierpower}
\end{eqnarray}

\subsection{ Case  $\sigma=1$ : Logarithmic barrier  }

For $\sigma=1$, Eq. \ref{teqsimplefinalexpli} becomes in terms of the length $L_N=2^N$
 \begin{eqnarray}
\sigma=1 \ \ : \ \ 
\ln \left[ 2^N t^{simple}_{eq}(L_N=2^N) \right] && 
% \simeq \beta J_0 \left[ 1 + \frac{4}{3} (N-1)+ \frac{1}{9} \right]
 =\beta J_0 \left[  \frac{4}{3} N - \frac{2}{9}    \right] +O( L_N^{-1})
\nonumber \\
&& = \beta J_0 \left[  \frac{4}{3 \ln 2 } \ln (L_N)   - \frac{2}{9}      \right] +O( L_N^{-1})
\label{teqsimplefinalsigmaeq1}
\end{eqnarray}
i.e. the energy barrier near zero temperature grows logarithmically with the length $L_N=2^N$
 \begin{eqnarray}
\frac{ \ln \left[2^N  t^{simple}_{eq}(L_N=2^N) \right] }{\beta} \simeq 
  \frac{4 J_0 }{3 \ln 2 } \ln (L_N)
\label{teqsimplefinalbarrierlog}
\end{eqnarray}

\subsection{ Generalization to other dynamics where $G_0(B)$ does not depend explicitly on $\beta$ }

\label{genesimple}

It is clear from Eq. \ref{teqfinal} that for all dynamics where $G_0(B)$ does not depend explicitly on $\beta$,
the equilibrium time will be dominated again by $\Sigma_{min}$ near zero temperature,
with the result (using Eq \ref{calBsoluzero})
 \begin{eqnarray}
t_{eq}(2^N)  && 
\simeq \frac{e^{  \beta \sum_{n=0}^{N-1} J_n } }{2^{N}}  
 \frac{  e^{- \beta \Sigma_{min}   }}
{  G_{0}\left( {\cal B}_0(\epsilon=-1,..,\epsilon_{N-1}=-1) \right)   }
\nonumber \\
&& \simeq \frac{e^{  \beta \sum_{n=0}^{N-1} J_n } }{2^{N}}  
 \frac{  e^{- \beta \Sigma_{min}   }}
{  G_{0}\left( J_{0} + \sum_{m=1}^{N-1} J_m \left( - \frac{1}{2} \right)^m \right)   }
\label{teqfinalothersimple}
\end{eqnarray}
i.e.  the dynamical barriers are the same as above in Eqs \ref{teqsimplefinalbarrierpower}
and Eq. \ref{teqsimplefinalbarrierlog} (the only difference is in the prefactor of the exponential).

\section{ Equilibrium time for the Glauber dynamics }

\label{sec_glauber}

\subsection{ Real Space Renormalization Solution }

For the initial condition of Eq. \ref{defglauber} describing the Glauber dynamics,
 Eq \ref{teqfinal} becomes
 \begin{eqnarray}
t^{Glauber}_{eq}(2^N) = \frac{1}{ 2 G^{final}_{N} } && 
= \frac{e^{  \beta \sum_{n=0}^{N-1} J_n } }{2^{N}}  
\sum_{\epsilon_{N-1}=\pm}  \sum_{\epsilon_{N-2}=\pm}  .. \sum_{\epsilon_1=\pm} 
  e^{- \beta \Sigma (\epsilon_1,..,\epsilon_{N-1})}  
\left[  e^{ \beta {\cal B}_0(\epsilon_1,..,\epsilon_{N-1})  }+  e^{ - \beta {\cal B}_0(\epsilon_1,..,\epsilon_{N-1})  }  \right]
\label{teqglauber}
\end{eqnarray}

\subsection{ Leading term near zero temperature }

So at low temperature, we have to minimize over the $\epsilon_i=\pm 1$ and over $\epsilon=\pm1$ 
the combination (Eqs \ref{calBsumSexpli} and Eq \ref{calBsoluzero})
 \begin{eqnarray}
\Sigma(\epsilon_1,..,\epsilon_{N-1}) +\epsilon {\cal B}_{0}(\epsilon_1,..,\epsilon_{N-1}) && =\sum_{m=1}^{N-1} J_m \left[ \sum_{n=1}^m \prod_{k=n}^m \left( \frac{\epsilon_{k}}{2} \right) \right]
+\epsilon \left[ J_{0} + \sum_{m=1}^{N-1} J_m \prod_{k=1}^m \left( \frac{\epsilon_{k}}{2} \right) \right]
\nonumber \\ &&
= J_0 \epsilon +
 J_1   \frac{\epsilon_{1}}{2} (1+\epsilon)
+  J_2 \frac{\epsilon_{2}}{2}  \left[ 1+ \frac{\epsilon_{1}}{2} (1+\epsilon)  \right]
+  J_3 \frac{\epsilon_{3}}{2}  \left[ 1+ \frac{\epsilon_{2}}{2} + \frac{\epsilon_{1} \epsilon_{2}}{2^2} (1+\epsilon) \right] +...
\nonumber \\ &&
+  J_{N-2} \frac{\epsilon_{N-2}}{2}
 \left[ 1+ \frac{\epsilon_{N-3}}{2} + \frac{\epsilon_{N-3} \epsilon_{N-4}}{2^2} +...
+ \frac{\epsilon_{N-3} \epsilon_{N-4} ... \epsilon_1 }{2^{N-3}} (1+\epsilon) \right]
\nonumber \\ &&
+  J_{N-1} \frac{\epsilon_{N-1}}{2}
 \left[ 1+ \frac{\epsilon_{N-2}}{2} + \frac{\epsilon_{N-2} \epsilon_{N-3}}{2^2} +...
+ \frac{\epsilon_{N-2} \epsilon_{N-3} ... \epsilon_1 }{2^{N-2}}(1+\epsilon)  \right]
\label{sigmaB}
\end{eqnarray}
The new factors containing $\epsilon$ are not able to change the signs of the expressions between brackets,
so we find again by recurrence (Eqs \ref{epslast} and \ref{epslastnext})
that this function is minimum by choosing all $\epsilon_i=-1$ and finally $\epsilon=-1$.
The corresponding minimum reads
 \begin{eqnarray}
\left[ \Sigma(\epsilon_1,..,\epsilon_{N-1}) +\epsilon {\cal B}_{0}(\epsilon_1,..,\epsilon_{N-1}) \right]_{min} && =  \Sigma(\epsilon_1=-1,..,\epsilon_{N-1}=-1) - {\cal B}_{0}(\epsilon_1=-1,..,\epsilon_{N-1}=-1)
\nonumber \\ &&
=\sum_{m=1}^{N-1} J_m \left[ \sum_{n=1}^m  \left( - \frac{1}{2} \right)^{m-n+1} \right]
- \left[ J_{0} + \sum_{m=1}^{N-1} J_m  \left( - \frac{1}{2} \right)^m \right]
\nonumber \\ &&
= - J_0 + \sum_{m=1}^{N-1} J_m \left[ \sum_{k=1}^{m-1}  \left( - \frac{1}{2} \right)^{k} \right]
\nonumber \\ &&
= - J_0 - \frac{1}{3}  \sum_{m=1}^{N-1} J_m  \left[ 1-  \left( - \frac{1}{2} \right)^{m-1} \right]
\label{sigmaBmin}
\end{eqnarray}

We thus obtain the following leading behavior at low temperature (Eq \ref{teqglauber})
 \begin{eqnarray}
t^{Glauber}_{eq}(2^N) && 
\simeq \frac{1 }{2^{N}} \  e^{  \beta \left[ \sum_{m=0}^{N-1} J_m - \left[ \Sigma(\epsilon_1,..,\epsilon_{N-1}) +\epsilon {\cal B}_{0}(\epsilon_1,..,\epsilon_{N-1}) \right]_{min} \right]}
\nonumber \\ && 
= \frac{1 }{2^{N}} \  e^{  \beta \left[ \sum_{m=0}^{N-1} J_m
  + J_0 + \frac{1}{3}  \sum_{m=1}^{N-1} J_m  \left[ 1-  \left( - \frac{1}{2} \right)^{m-1} \right] \right]}
\nonumber \\ && 
= \frac{1 }{2^{N}} \  e^{  \beta \left[ 2 J_0 + \sum_{m=1}^{N-1} J_m
 \left( \frac{4}{3}- \frac{1}{3} \left( - \frac{1}{2} \right)^{m-1} \right) \right] }
\label{teqglauberfinal}
\end{eqnarray}

We may now use Eq \ref{Jn} to compute explicitly
 \begin{eqnarray}
\ln \left[ 2^N t^{Glauber}_{eq}(2^N) \right] && 
=    \beta \left[ 2 J_0 + \sum_{m=1}^{N-1} J_0 2^{(1-\sigma)m }
 \left( \frac{4}{3}- \frac{1}{3} \left( - \frac{1}{2} \right)^{m-1} \right) \right]
\nonumber \\ && 
=    \beta J_0 \left[ 2  + \frac{4}{3}\sum_{m=1}^{N-1}  2^{(1-\sigma)m }
  - \frac{1}{3} \sum_{m=1}^{N-1}  2^{(1-\sigma)m }
 \left( - \frac{1}{2} \right)^{m-1} 
 \right]
\nonumber \\ && 
=    \beta J_0 \left[ 2  + \frac{4}{3} 2^{1-\sigma}\frac{1-   2^{(1-\sigma)(N-1)} }{1-2^{1-\sigma} }
  - \frac{1}{3}  2^{1-\sigma}   \frac{1-   (-2^{-\sigma})^{(N-1)} }{1+2^{-\sigma} }
 \right]
\label{teqglauberfinalexpli}
\end{eqnarray}
This expression is valid for finite $N$ and we may check the first terms
 \begin{eqnarray}
\ln \left[ 2 t^{Glauber}_{eq}(2) \right] &&   = 2 \beta J_0
\nonumber \\
\ln \left[ 2^2 t^{Glauber}_{eq}(2^2) \right]  &&  =   \beta J_0 \left[ 2  +  2^{1-\sigma}
 \right] = \beta \left[ 2 J_0 +  J_1
 \right]
\label{teqglauberfirst}
\end{eqnarray}

To obtain the leading behavior for large $N$, we should now specify the value of $\sigma$.

\subsection{ Case  $\sigma<1$ : Power-law barrier  }

 For $\sigma<1$, Eq \ref{teqglauberfinalexpli} becomes in terms of the system size $L_N=2^N$
 \begin{eqnarray}
\sigma<1 \ \ : \ \ 
\ln \left[ 2^N t^{Glauber}_{eq}(2^N) \right] && 
\simeq   \beta J_0 \left[
\frac{4}{3} \frac{   2^{(1-\sigma) N} }{2^{1-\sigma}-1 }
+  2  + \frac{4}{3} \frac{1 }{2^{\sigma-1}-1 }
  - \frac{2}{3}    \frac{1 }{2^{\sigma}+1 } \right]
\nonumber \\
&& = \beta J_0 \left[
\frac{4}{3} \frac{   L_N^{(1-\sigma) } }{2^{1-\sigma}-1 }
+  2  + \frac{4}{3} \frac{1 }{2^{\sigma-1}-1 }
  - \frac{2}{3}    \frac{1 }{2^{\sigma}+1 } \right]
\label{teqglauberfinalsigmal1}
\end{eqnarray}
i.e. the energy barrier near zero temperature scales as the same power-law of the length $L_N=2^N$
 \begin{eqnarray}
\frac{ \ln \left[ 2^N  t^{Glauber}_{eq}(L_N=2^N) \right] }{\beta} \simeq
  \frac{ 4 J_0 }{3( 2^{1-\sigma}-1)}  L_N^{(1-\sigma)} 
\label{teqglauberfinalbarrierpower}
\end{eqnarray}
and with the same prefactor as in Eq. \ref{teqsimplefinalbarrierpower},
even if the finite corrections of Eq \ref{teqglauberfinalsigmal1} are different
the finite corrections of Eq \ref{teqsimplefinalsigmal1}
(more explanations on these finite differences are given in Appendix \ref{sec_barrier}).

\subsection{ Case  $\sigma=1$ : Logarithmic barrier  }

 For $\sigma=1$, Eq \ref{teqglauberfinalexpli} becomes in terms of the system size $L_N=2^N$
 \begin{eqnarray}
\sigma=1 \ \ : \ \ 
\ln \left[ 2^N t^{Glauber}_{eq}(2^N) \right] 
&& \simeq   \beta J_0 \left[ 2  + \frac{4}{3} (N-1)  - \frac{2}{9}     \right]
=  \beta J_0 \left[  \frac{4}{3} N  + \frac{4}{9}     \right]
\nonumber \\
= \beta J_0 \left[  \frac{4}{3 \ln 2 } \ln L_{N}  + \frac{4}{9}     \right]
\label{teqglauberfinalsigmaeq1}
\end{eqnarray}
i.e. the energy barrier near zero temperature grows with the same logarithmic barrier
 \begin{eqnarray}
\frac{ \ln \left[ 2^N t^{Glauber}_{eq}(L_N=2^N) \right] }{\beta} \simeq 
  \frac{4 J_0 }{3 \ln 2 } \ln (L_N)
\label{teqglauberfinalbarrierlog}
\end{eqnarray}
with the same prefactor as in Eq \ref{teqsimplefinalbarrierlog}, 
even if the finite corrections of Eq. \ref{teqsimplefinalsigmaeq1} and of Eq. \ref{teqglauberfinalsigmaeq1}
are different (see more details in Appendix \ref{sec_barrier}).

\subsection{ Generalization to other dynamics where $G_0(B) \propto e^{- \beta \vert B \vert }$  }

\label{geneglauber}

From the analysis presented above, it is clear that dynamical barriers near zero temperature
will remain the same for all dynamics where the amplitude $G_0(B)$ displays
 the same exponential decay as the Glauber amplitude of Eq. \ref{defglauber}
 \begin{eqnarray}
G_0(B) \propto e^{- \beta \vert B \vert }
\label{G0decay}
\end{eqnarray}
As an example of other dynamics with the same exponential decay, we may cite the Metropolis dynamics corresponding to
(see Eq \ref{wgene}) 
\begin{eqnarray}
G^{Metropolis}_0(B) = {\rm min} ( e^{\beta B}, e^{- \beta B}  ) = e^{- \beta \vert B \vert }
\label{G0metropolis}
\end{eqnarray}

\section{ Conclusion}

 \label{sec_conclusion}

In this paper, we have studied the stochastic dynamics of the Dyson hierarchical one-dimensional Ising model
 of parameter $0<\sigma \leq 1$ via the
Real Space Renormalization introduced in our previous work \cite{us_rgdyna}.
We have shown that this renormalization procedure amounts to renormalize
 a single function $G(B)$
that defines the transition rates of the renormalized dynamics.
We have solved explicitly the RG flow for two types of dynamics, namely the 'simple' dynamics
(and other equivalent dynamics of section \ref{genesimple})
 and the Glauber dynamics (and other equivalent dynamics of section \ref{geneglauber}).
We have obtained that the leading diverging dynamical barrier is the same power-law 
$\ln t_{eq}(L) \simeq \beta \left( \frac{ 4 J_0 }{3( 2^{1-\sigma}-1)} \right)  L^{1-\sigma}$ for $\sigma<1$
and the same logarithmic term :  $\ln t_{eq}(L) \simeq \left[ \beta  \left( \frac{ 4 J_0 }{3 \ln 2} \right) -1 \right] \ln L $
for $\sigma=1$, even if finite corrections are different, as explained in Appendix \ref{sec_barrier}.

\appendix

\section{ Link between dynamical barriers and the highest energy cost of a domain-wall }

\label{sec_barrier}

\subsection{ Case of the Glauber dynamics }

For the Glauber dynamics, the dynamical barrier for a system of $2^N$ spins corresponds to the  
maximal energy cost of one domain wall inside the system
 \begin{eqnarray}
\frac{1}{\beta} \ln \left[ 2^N t^{Glauber}_{eq}(2^N) \right]
 &&  = {\max \limits_{0 \leq k \leq 2^N} } \ \left(U_{2^N}^{(k,2^N-k)} -  U_{2^N}^{GS} \right) 
\label{linkbarrierworst}
\end{eqnarray}
where $U_{2^N}^{(k,2^N-k)} $ represents the energy of the configuration where the 
 first $k$ spins are $(-1)$, whereas all others spins are $(+1)$.

The property of Eq. \ref{linkbarrierworst} can be easily checked for small systems, for instance for the case $N=1$ containing $2^1=2$ and the case $N=2$ containing $2^2=4$  spins (Eq \ref{teqglauberfirst})
 \begin{eqnarray}
\frac{1}{\beta} \ln \left[ 2 t^{Glauber}_{eq}(2) \right] &&   =  2 J_0 = U_{2}^{(1,1)} -  U_{2}^{GS}
\nonumber \\
\frac{1}{\beta} \ln \left[ 2^2 t^{Glauber}_{eq}(2^2) \right]  &&  =   2 J_0 +  J_1 = U_{4}^{(1,3)} -  U_{4}^{GS} 
= U_{4}^{(3,1)} -  U_{4}^{GS}
\label{bteqglauberfirst}
\end{eqnarray}

For an arbitrary number $N$ of generations containing $2^N$ spins,
 the correspondence of Eq. \ref{linkbarrierworst}
is less straightforward because 
the energy cost of the configuration where the 
 first $k$ spins are $(-1)$, whereas all others spins are $(+1)$ reads
\begin{eqnarray}
U_{2^N}^{(k,2^N-k)} -  U_{2^N}^{GS}  && = 2 J_0  c_0(k) + 
\sum_{j=1}^{N-1} J_j \left[ 2 c_j(k) + \left(1- 2 c_j(k) \right) 2^{1-j} \Sigma_{j-1}(k) \right]
\label{Udomainkbinary}
\end{eqnarray}
in terms of the coefficients $c_i(k) \in \{0,1\}$ of the base-two decomposition 
\begin{eqnarray}
k= \sum_{i=0}^{N-1} c_i(k) 2^i = c_0(k)+c_1(k) 2 + c_2(k) 2^2 +... 
\label{kbinary}
\end{eqnarray}
and of the corresponding partial sums for $i=0,..,N-1$
\begin{eqnarray}
\Sigma_i(k)= \sum_{j=0}^{i} c_j(k) 2^j 
\label{kbinarypartial}
\end{eqnarray}
Nevertheless, one can check that the energy barrier obtained in Eq. \ref{teqglauberfinal}
corresponds to the energy cost of a domain wall located at $k_*$
 \begin{eqnarray}
\frac{1}{\beta} \ln \left( 2^N t^{Glauber}_{eq}(2^N) \right) 
&& \simeq 2 J_0 + \sum_{m=1}^{N-1} J_m \left[ \frac{4}{3}- \frac{1}{3} \left( - \frac{1}{2} \right)^{m-1} \right] 
= \left(U_{2^N}^{(k_*,2^N-k_*)} -  U_{2^N}^{GS} \right) 
\label{bteqglauberfinal}
\end{eqnarray}
where $k_*$ is characterized by the following coefficients in the base-two decomposition \ref{kbinary}
\begin{eqnarray}
c_{2p}(k_*) && = 1 \nonumber \\
c_{2p+1}(k_*) && = 0
\label{kbinarymax}
\end{eqnarray}
so that
\begin{eqnarray}
k_*= 1+2^2+2^4+..
\label{kbinarystar}
\end{eqnarray}

\subsection{ Case of the simple dynamics  }

For the 'simple' dynamics, the correspondence of Eq. \ref{linkbarrierworst}
between the dynamical barrier and the maximal energy cost of a single domain wall 
does not hold,
as can be seen already for 
the case $N=2$ corresponding to $2^1=2$ spins and 
for the case $N=2$ corresponding to $2^2=4$ spins, 
since the dynamical barriers of Eq \ref{teqsimplefirst}
 \begin{eqnarray}
\frac{1}{\beta} \ln \left[ 2 t^{simple}_{eq}(2) \right] &&   =  J_0
\nonumber \\
\frac{1}{\beta} \ln \left[ 2^2 t^{simple}_{eq}(2^2) \right]  && = J_0+\frac{3}{2} J_1 
\label{bteqsimplefirst}
\end{eqnarray}
are clearly different from the maximal energy cost of Eq \ref{bteqglauberfirst}.
This can be understood as follows on the case $N=1$ with two spins.
The transitions rates associated to the simple dynamics 
(Eqs \ref {wgene} and \ref{defsimple})
 read
 \begin{eqnarray}
W^{simple}(++ \to +-) && =W^{simple}(++ \to -+) = e^{- \beta J_0} 
\nonumber \\
W^{simple}( +- \to ++ ) && =W^{simple}(+- \to --) = e^{+ \beta J_0} 
\label{rates2spinssimple}
\end{eqnarray}
whereas the Glauber transition rates (Eq \ref{defglauber}) are given by
 \begin{eqnarray}
W^{Glauber}(++ \to +-) && =W^{Glauber}(++ \to -+) = \frac{e^{- \beta J_0} }{e^{+ \beta J_0}+e^{- \beta J_0} }
 = \frac{e^{- 2 \beta J_0} }{1+e^{- 2 \beta J_0} }
\nonumber \\
W^{Glauber}( +- \to ++ ) && =W^{Glauber}(+- \to --) =  \frac{e^{+ \beta J_0} }{e^{+ \beta J_0}+e^{- \beta J_0} }
 = \frac{1 }{1+e^{- 2 \beta J_0} }
\label{rates2spinsglauber}
\end{eqnarray}
For $2$ spins, the equilibrium time is determined by the rate $W(++ \to +-)$
to create a domain-wall when starting from one ground state
(the time to eliminate the domain-wall is then negligible), and these two rates are 
respectively of order $ e^{- \beta J_0}  $ for the simple dynamics and of order $e^{- 2 \beta J_0}$ for the Glauber dynamics,
i.e. the dynamical barriers differ by a factor $2$.
One could argue that the Glauber dynamics (or other equivalent dynamics of section \ref{geneglauber}) is more 'physical', in the sense that all transitions rates remain bounded
near zero-temperature, whereas in the 'simple' dynamics
(or other equivalent dynamics of section \ref{genesimple}), transition rates corresponding to a decrease of the energy
diverge near zero temperature.
Nevertheless, we should stress that
 the difference between the dynamical barriers of the two dynamics
remains of order $O(1)$, whereas the leading terms found in the text for the case $\sigma<1$
(Eqs \ref{teqsimplefinalbarrierpower} and \ref{teqglauberfinalbarrierpower})
and for the case $\sigma=1$ (Eqs \ref{teqsimplefinalbarrierlog} and \ref{teqglauberfinalbarrierlog}) are the same.

\end{document}